\documentclass{iauc}
\usepackage{graphicx}

\title[The environment of dwarf spheroidal satellites] 
{The environment of dwarf spheroidal satellites; ram pressure, tides 
and external radiation fields}

\author[Lucio Mayer]   
{Lucio Mayer$^1$%
,
email:lucio@physik.unizh.ch\\
}
\affiliation{$^1$ Institute for Theoretical Physics,
University of Z\"urich, Winterthurerstrasse 190, 8057 Zurich, Switzerland
}\break 
\pubyear{2005}
\volume{198} 
\pagerange{119--126}
\date{?? and in revised form ??}
\jname{Proceedings Title IAU Colloquium}
\editors{H. Jerjen and B.Binggeli, eds.}
\begin{document}

\maketitle

\begin{abstract}
We discuss the role of environmental mechanisms in the evolution of dwarf
galaxy satellites using high-resolution N-Body+SPH simulations that include
simultaneously tidal forces, ram pressure and heating from ionizing radiation
fields. Tidally induced bar-buckling instabilities can transform a rotating 
disky dwarf into a pressure supported spheroidal. Efficient gas removal  
requires instead a combination of tidal mass loss and ram pressure stripping
in a diffuse gaseous corona around the primary system. The efficiency of ram
pressure depends strongly on how extended the gas remains during the
evolution. Bar driven inflows that tend to drive  the gas to the bottom of the
potential well can be opposed by the heating from external radiation
fields. We show that 
even fairly massive dwarfs ($V_{peak} >$ 30 km/s) would be  
stripped of their gas over a few Gyr if they enter the Milky Way halo at
$z > 2$ thanks to the effect of the cosmic UV background. Gas mass loss
can be much faster, occurring in less than 1 Gyr, if dwarf satellites
have their first close approach with the primary at the epoch of bulge
formation.  Indeed at that time the primary galaxy should have a FUV
luminosity comparable to that of major present-day starbursts, resulting in 
a local UV field even more intense than the cosmic background.

\end{abstract}

\section{Introduction}

Dwarf spheroidals (dSphs) are the faintest galaxies known. They are gas poor
and have pressure supported stellar components (Mateo 1998). Among them some
stopped forming stars about 10 Gyr ago and other have extended star formation
histories (Hernandez et al. 2000). 
They are typically clustered around the largest galaxy in a
group. Both mass loss from supernovae winds (Dekel \& Silk 1986) and 
environmental mechanisms like tidal and ram pressure stipping 
(Einasto et al. 1974) 
have been invoked to explain their properties. Suppression of gas 
accretion and/or photoevaporation during the reionization epoch likely played
a role as well (Bullock et al. 2000).
In this paper we describe the results of N-Body/SPH simulations of disky 
dwarf galaxies subject to the combined action of the main environmental
mechanisms, tides, ram pressure, and both a cosmic and local ultraviolet
radiation field. The model galaxies and the choice of the orbits are
consistent with the predictions of $\Lambda$CDM models 
(Mayer et al. 2002, 2005,hereafter MA05).
These simulations allow for the first time to explore the overall effect 
of the environment.

\section{The masses of dwarf spheroidals}

Knowing the present and past mass of dSphs is crucial in order
to compute the effects of both environmental and internal mechanisms that
might affect their evolution. Moreover, the mass of dSphs has
important implications on the missing satellites problem (Moore et al. 1999),
namely the fact that CDM predicts an order of magnitude more satellites
than observed around the Milky Way or M31.
S.D.M. White (2000) has suggested that this problem could be alleviated if 
the peak circular velocities of the halos of dSphs, and thus their
masses, are higher than 
those obtained from their central velocity dispersion under the
assumption of a flat (isothermal) rotation curve. This is expected 
if dSphs sit at the center of an extended halo with an non-isothermal
(e.g. NFW) profile, thus having circular velocity profiles that keep
rising well beyond the apparent tidal radii of the galaxies.
Stoehr et al. (2002) fitted the kinematics of dSphs inside subhaloes
of a cosmological $\Lambda$CDM simulation using Jeans modeling and 
King profiles for the light distribution of dSphs
and found that these galaxies live in halos having a peak circular
velocity $V_{peak} = 40-60$ km/s. The halo density profiles employed in
their analysis are much flatter than an NFW profile at the scale of the core 
radius of dSphs.  

\begin{figure}
\hskip 2truecm
\includegraphics[height=3.5in,width=3.5in,angle=0]{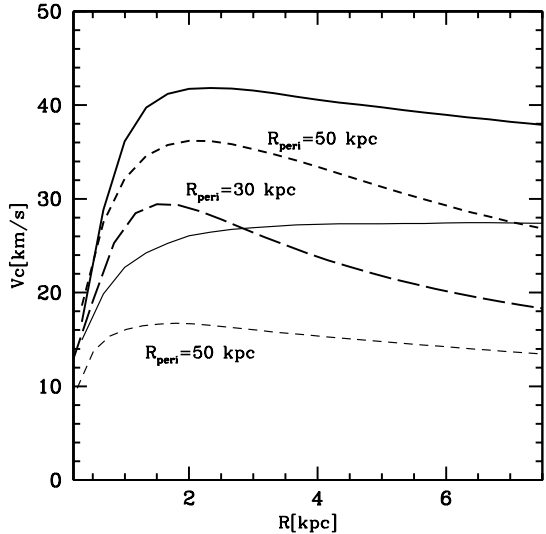}
  \caption{Evolution of the rotation curves of dwarf models with
$V_{peak}=40$ km/s (thick lines)and $V_{peak}=28$ km/s (thin lines). The
solid lines represent the initial conditions while the short-dashed show
the curves after 3 orbits. The thick long-dashed line shows the
curve for the same dwarf model with $V_{peak}=40$ km/s on a different
orbit. Pericenter distances for the various orbits are indicated in the 
plot (see MA05).}
\end{figure}

Kazantzidis et al. (2004) have studied the tidal disruption of individual 
CDM satellites with as many as $10^7$ particles and of
subhaloes in a hi-res $\Lambda$CDM simulation. They repeated the analysis of Stoehr
et al. (2002) and found that the observed velocity dispersions of Draco
and Fornax can only be reproduced in subhaloes with $V_{peak} =20-30$ km/s. 
In their simulations halos maintain
cuspy profiles down to the force resolution, which is
smaller than the core radius of dSphs ($\sim 100$ pc). As a result
the rotation curves rise more steeply than those of Stoehr et al. (2002),
explaining the discrepancy.
The masses of the tidally truncated subhaloes of Kazantzidis et al. (2004)
are in the range $10^8-10^9 M_{\odot}$, in agreement with
more recent analysis based on newer data on stellar velocities (e.g. 
Wilkinson et al. (2001)). These masses are still
quite high, yielding $M/L > 100$ for Draco. However peak 
circular velocities are at most a factor of 2 higher than those used
in Moore et al. (1999), with the consequence that little changes for the
missing satellite problem.

The initial $V_{peak}$  of dSphs was larger than 30
km/s since mass loss produced by tidal shocks can lower $V_{peak}$
(Ghigna et al. 1998). Kravtsov et al. (2004)), using hi-res cosmological
simulations without baryons, find that $V_{peak}$ can drop by a
factor of 2 in 10 Gyr or so. Mayer et al. (2002) and MA05 always find a smaller 
decrease in $V_{peak}$ in their large set of SPH simulations that follow
the evolution of satellites with baryonic disks embedded in CDM haloes
(Figure 1).
The different results might be due to the fact that some of the satellites
in Kravtsov et al. (2004) have orbits with pericenters smaller than those
considered by MA05. However, the inclusion of baryons
also makes the satellites more robust to the tidal disruption of the inner
region, which is what matters for the evolution of $V_{peak}$. Whether
or not the baryonic disk becomes bar unstable also is important, 
since bar-driven
inflows can deepen the central potential of the dwarf (Kazantzidis, Mayer
et al., in preparation). High-resolution cosmological simulations with
hydrodynamics will eventually be able to quantify the overall
significance of these effects. In any case, the results of MA05 
suggest that the initial $V_{peak}$ of dSphs was in the 
range $35-50$ km/s. This has important implications. First, it means
that photoevaporation did not play a major role during the early evolution
of satellites at high $z$ since it is only effective for $V_{peak} <
20$ km/s (Susa \& Umemura 2004; Shaviv \& Dekel 2003). This is consistent
with the lack of a clear signature of the reionization epoch in the star 
formation histories of dSphs (Grebel \& Gallagher 2004).
Second, blow-out of the gas of dSphs due to supernovae winds 
cannot have occurred
since it requires $V_{peak} < 30$ km/s (e.g. MacLow \& Ferrara 1999).

\begin{figure}
\hskip 2truecm
 \includegraphics[height=3.2in,width=3.2in,angle=0]{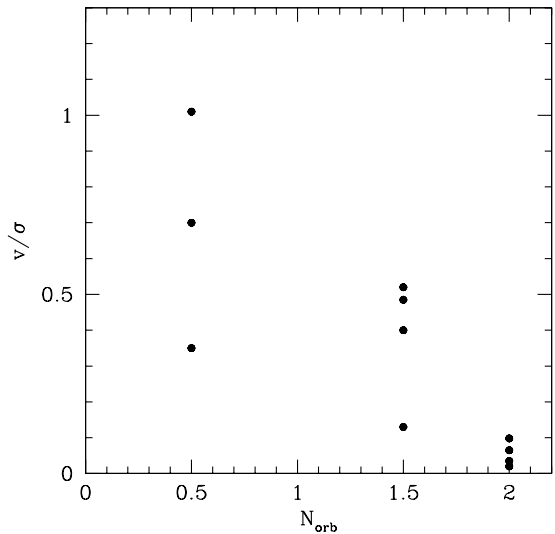}
 \caption{$v/\sigma$ as a function of the number of orbits for the satellites 
in the cosmological hydrodynamical simulation of Governato, Mayer et al. 
(2004).$v/\sigma$ is measured within the half mass radius of the dwarfs.
The mean number of orbits for each of the three groups of dwarfs appears on the
horizontal axis.}
\end{figure}

\section{Morphological evolution of disky dwarfs into dSphs; tidal stirring}

Aside from the different gas content, three other facts
have to be considered when comparing dSphs and gas-rich, similarly faint
dwarf galaxies known as dwarf irregulars (dIrrs).
First, except for the faintest among them all dIrrs in the Local
Group and nearby groups exhibit substantial rotation while dSphs
do not. Second, stellar profiles in the two classes of galaxies are similarly 
close to exponential. Third, a morphology density relation exists such that
dSphs are clustered around the primary galaxies while dIrrs are found at 
much larger distances from them.
Interestingly, Karachentsev (these proceedings)
finds that the fraction of dSphs over the total number of dwarfs in several 
nearby groups (including the Local Group)
decreases drastically at distances larger than about $250$ kpc from the
primary galaxies, these being comparable to the virial radius of haloes
hosting bright spiral galaxies in $\Lambda$CDM models. This clearly suggests
that only dSphs are bound satellites of the primaries and thus the
environment must be playing a crucial role in differentiating dIrrs from
dSphs. Of course there are outliers like the Local Group 
dSphs Cetus and Tucana,
located at more than 500 kpc from, respectively, M31 and the MW, but
this is seen also in cosmological simulations, where a few satellites
on very plunging orbits can have apocenters exceeding the
virial radius of the primary (Ghigna et al. 1998). 

Mayer et al. (2001a,b) and MA05 have shown that repeated tidal shocks at
pericenters of their orbits within the halo of a massive spirals can 
transform disky dwarfs into objects resembling dSphs. The timescale
of the transformation is a few orbital times (several Gyr). The 
mechanism behind the transformation has to do with non-axisymmetric 
instabilities of stellar disks. First, tidal shocks induce strong bar
instabilities in otherwise stable, light disks resembling those of 
present-day dIrrs. Second, the bar buckles due to the amplification
of vertical bending modes and turns into a spheroidal component. Finally, 
tidal heating/tidal mass loss thicken/remove the disk outside the buckled
bar. We stress that stellar mass loss can be 
minimal when the disk sits deep in the potential well of a massive halo
(see the GR8 run in Mayer et al. 2001b) and yet the transformation occurs.
The remnants have nearly exponential profiles; the brightest among them, 
those coming from progenitors
having relatively massive disks ($ > 10^8 M_{\odot}$), develop a central
steepening of the profile due to the particularly strong bar, and overall their
profile resembles that of the bright dwarf elliptical satellites of
M31, like NGC205.
Most importantly, the bar sheds angular momentum outwards and as a result
the systems end up with low rotation, with a typical final 
$v/\sigma < 0.5$.  The transformation of course needs the dwarf to be
on a bound orbit; for the mean apocenter/pericenter ratio found in
cosmological simulations, $\sim 5$, Mayer et al. (2001b) determine that
an apocenter distance comparable to the distance of Leo I (250 kpc), the
farthest dSph satellite of the MW, is a limiting case for producing a
dSph from a disky dwarf in less than 10 Gyr (Tucana and Cetus should thus
be on radial orbits).

In brief this ``tidal stirring'' accounts for most of the similarities
and differences between dIrrs and dSphs, including the existence of the
morphology density relation, by postulating progenitors of dSphs with
light, low surface brightness disks embedded in massive halos 
like dIrrs. This does not mean that such progenitors were identical
to present-day dIrrs. Since they formed at high
$z$ they likely had assembly histories quite different from present-day
dIrrs, and thus their stellar populations and metallicities were 
probably different. 
The missing piece now is how to place this model in the context of
hierarchical structure formation. Unfortunately, to date cosmological 
simulations with hydrodynamics have not allowed a robust analysis of the
structural evolution of satellites due to their limited resolution
(typically the force resolution is 500-1 kpc, so the tiniest dSphs like
Draco are not resolved at all). 
Searching for evidence of tidal stirring in these
simulations is thus quite hard, but it can be done for a 
few well resolved dwarfs. We selected the brightest satellites 
($M_B > - 14$) of the large spiral galaxy in the cosmological hydro 
simulation of Governato, Mayer et al. (2004, hereafter GM04) as well as similarly bright
dwarfs outside the virial radius of the same galaxy. Figure 2
shows their $v/\sigma$ within the effective radius versus 
the number of orbits performed within the main system. 
Clearly $v/\sigma$ correlates
well with the number of orbits as expected within the tidal stirring
scenario. We also found that the shape of the stellar components of the
dwarfs goes from more disky to more spheroidal with increasing number
of orbits. The absolute values of the $v/\sigma$ have to be taken with caution
because numerical two-body heating (e.g. Mayer 2004) is certainly an 
issue for these objects (halos have only a few thousand particles at 
this scale). However the trend is evident.

\begin{figure}
\hskip 2.2truecm
\includegraphics[height=3.4 in,width=3.4 in,angle=0]{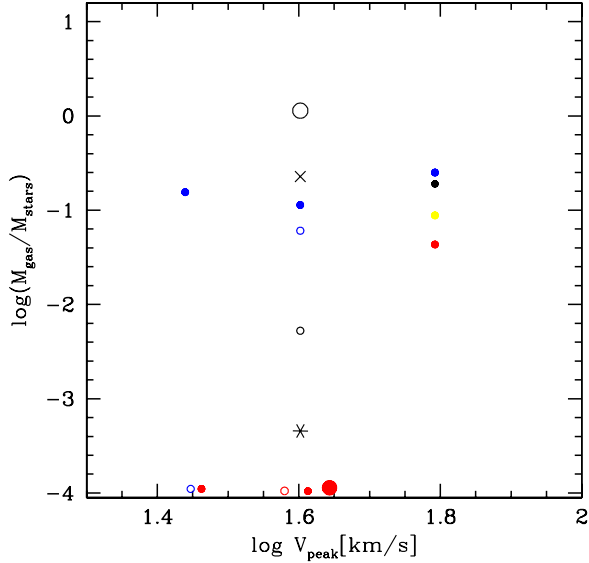}
  \caption{Gas mass fraction (relative to the stellar mass) of simulated
dwarf galaxy satellites  plotted againts 
the initial peak velocity of the model galaxies.
Data are taken from the
N-Body/SPH simulations described in MA05 plus new
runs employing a local, time-dependent UV radiation field (see section
4). Measurements
are done after 2 orbits within the Milky Way halo. The lower limit on the
vertical axis corresponds to the resolution limit in the simulations
(therefore points at the very bottom correspond to zero gas mass as measured 
in the simulation). Some symbols have been displaced along
the horizontal axis to avoid overlaps. Filled symbols refer to runs with
pericenters of 50 kpc, open symbols refer to runs with pericenters of 30
kpc. Large symbols are for galaxies with a disk gas mass fraction of $0.9$,
small symbols are for those with a disk gas mass fraction of $0.3$ (see
section 4). Red is used for adiabatic runs, blue for runs with radiative
cooling, black for runs with radiative cooling+cosmic UV background and yellow
for runs with cooling, UV and star formation. 
The cross and star symbols are used for two
runs with radiative cooling and a local UV radiation field yielding a maximum 
heating rate, respectively, 5 times and 10 times higher than that associated
with the cosmic background.}
\end{figure}

\section{Gas stripping; ram pressure, tides and ultraviolet radiation}

Mayer et al. (2001b) showed that tidal stripping alone could not
produce the low gas fractions found in dSphs starting from a gas-rich
disky dwarf. Gas consumption by star formation does not change the
results significantly for mean star formation rates consistent
with those inferred for dSphs (Hernandez et al. 2000).
MA05 study the combined effect of ram pressure and tidal
stripping. They construct two-component models for the Milky Way halo in
which a dark matter halo consistent with the results of $\Lambda$CDM 
simulations
has an embedded diffuse gaseous component with a temperature of $\sim
10^6$ K and a density of about $\sim 10^{-4}$ atoms/cm$^3$ at 30 kpc from
the center, consistent with the values inferred from OVI absorption
measurements and the existence of the Magellanic Stream 
(Sembach et al. 2003). Dwarf galaxies are placed on eccentric orbits
with pericenters of 30 or 50 kpc. 
In these simulations gas-rich disky dwarfs ($M_{gas}/M_{stars} \ge 0.4$) 
embedded in massive 
dark haloes typically lose 90\% of their gas content. 
In fact ram pressure increases by a factor up to 10 the amount of
stripped gas mass compared to the case in which only tides are included.
Ram pressure strongly depends on the depth of the
potential well of the dwarfs. While for dwarfs with $V_{peak} \le 30$ km/s
most of the gas content is easily removed, for more massive dwarfs 
the end result depends a lot on the orbit and on the temperature evolution
of their gas (see Figure 3).
The pericenter distance sets the strength of the
ram pressure force mostly through the dependence on the orbital speed of
the galaxy at pericenter. The temperature evolution determines whether
the gas component stays extended or becomes concentrated in the central,
deeper part of the potential well as a result of tidally induced
bar-driven inflows. Compression from the outer medium heats the gas. 
The gas is rapidly heated
to $10^5$ K, where the cooling function peaks (the initial temperature is
$\sim 8000$ K)
at which point it cools radiatively to $10^4$ K in a fraction of
a dynamical time; this cold gas easily sinks towards the center due
to the torque exerted by the bar and cannot be removed by ram pressure.
Instead, if compressional heating is
not radiated away, as when the gas evolves adiabatically, the increased 
pressure opposes the bar-driven inflow, keeping more gas at larger radii
where the it is easily stripped. 

For stripping
to be very effective for dwarfs with initial $V_{peak} \sim 40-50$ km/s it is
sufficient that the  temperature of the gas is kept above $10^4$ K (the
virial temperature of these halos is $3-5 \times 10^4$ K). This
requires some heating source to counteract radiative cooling. 
MA05 find that the (uniform) cosmic UV background at $z >2$
(Haardt \& Madau 1996) can achieve this. 
They find that if the progenitors of Draco or Ursa Minor fell
into the Milky Way at $z > 2$ then ram pressure combined with tides was
able to remove their entire gas content in a couple of orbits. This translates
into a timescale of about 2-3 Gyr for a reasonable orbital time 
(namely one consistent with the current orbital distance). In this
scenario the observed truncation of the star formation in Draco and Ursa
Minor more than 10 Gyr ago occurred  as a consequence of the infall of 
these galaxies into the Milky Way halo. 
Draco would have formed most of its stars before infall
since most of the gas becomes ionized while
approaching pericenter for the first time.

However, MA05 also find that the exact gas fraction of gas that becomes 
ionized and then stripped is quite sensitive on the initial gas density 
(Figure 3). Starting from 
initial conditions having an exponential disk comprising about $4 \%$ of the virial mass(comparable to disk mass fractions of dIrrs), a disk with $30\%$ gas
fraction becomes entirely ionized while one with $90\%$ gas fraction 
remains neutral in the center and is only marginally stripped. 
The mass of neutral gas could produce a stellar mass comparable
to the present-day luminosity of Draco assuming star formation
goes on for another few Gyr.
This would yield a star formation history more extended than
that of Draco, possibly closer to that of Carina and Fornax.
On the other end a progenitor with a mostly gaseous disk that undergoes
complete stripping is an attractive case since it would naturally
produce the large $M/L$ of Draco without the need of invoking an extremely
low disk mass fraction in the initial conditions (MA05).
One obvious solution is to make the case for a 
higher heating and photoionization
rate; no dramatic increase of the rates would be needed since the 
neutral gas has a temperature 
only 20\% lower than $10^4$ K in the gas rich models.
The local UV flux coming
from the primary galaxy during a phase of intense star formation could 
have been stronger than the average cosmic background.
Mashchenko et al. (2004) calculate
that the FUV flux of M31 derived from its $H_{\alpha}$ luminosity 
is higher than the present-day value of the cosmic UV background
out to 10-20 kpc from its center.
Current starbursts have FUV luminosities $10^3-10^4$ higher than the MW and 
M31, in the range $10^{44}-10^{45}$ erg/s (Leitherer et al. 2002).
According to cosmological simulations at $z=2-3$ the Milky Way was
hosting a major central starburst as the 
bulge was being assembled from a big merging event (GM04).
A starburst is also predicted by models of the star formation history in the 
galactic bulge, which yield peak 
star formation rates of up to $100 M_{\odot}$/yr
(Elmegreen 1999), comparable to the most spectacular among present-day starbursts.
Lower quantities of dust expected given the lower metallicity of the gas at 
high redshift imply an escape fraction of ultraviolet photons higher than
that in present-day starbursts. But let us be conservative and
just assume  that the FUV luminosity of the bulge
was comparable to that of a major present-day starburst ($L_{FUV}=10^{45}$
erg/s). Under this hypothesis one obtains that at 30 kpc from the bulge the
heating rate was 10 times higher than that associated 
with the metagalactic UV background at $z=2-3$.
We have run new simulations
including such a local, time-dependent UV background 
(the intensity of the flux is modulated by the orbital distance of the
dwarf) and found that ram 
pressure stripping at the first  pericenter passage can be greatly
enhanced for a gas-rich dwarf.
Even the most dark matter dominated among our models, with $V_{peak}=
$40 km/s, loses more than 90\% of its gas along the first orbit and the
remaining gas is completely ionized. The ionized gas is then 
stripped along the second orbit (Figure 3).
This simulation reproduces naturally
both the early truncation of the star formation in Draco and its present-day 
high $M/L$ ratio ($> 100$).

\section{Towards a coherent picture; from gas stripping and gas inflows 
to the star formation histories}

A coherent picture is emerging. Dwarf spheroidals live in tidally truncated 
halos with moderate masses, $10^8-10^9 M_{\odot}$, yet this implies $M/L > 100$ 
in some of them (e.g. Draco and Ursa Minor). However, since simulations show 
that most  of the dark matter halo was stripped as the dwarfs entered the
primary halos, their original dark matter content was higher, resulting in
an initial $V_{peak} > 30$ km/s.
In fact tidal stripping can remove mass even quite close to the
center, reducing their $V_{peak}$ by $30-50$\% (see Figure 1).
The tidal interaction with 
the massive halos of the primaries is also responsible for the morphological 
transformation of small disks into spheroidal systems. Gas stripping instead
is the result of the combination of tidal and ram pressure stripping.
However a heating source is required in order to keep the gas extended and 
allow efficient stripping for dwarfs with $V_{peak} > 30$ km/s. 
The more attractive scenario is one in which 
the heating was provided by the cosmic UV background at high redshift 
aided by an even stronger local FUV field produced by the primary galaxies 
during an epoch of intense star formation. According
to galaxy formation models this epoch should coincide with the time of
bulge formation, at $z \ge 2$. This would 
explain the  sudden truncation of star formation that occurred in Draco 
and Ursa Minor about 10 Gyr ago. 
Dwarfs that entered the Milky Way halo later would
have kept some gas for at least one or two orbits and thus undergo extended
star formation.  Our simulations show that the remaining gas falls towards the
center as a result of bar driven inflows at pericenter passages. The central
gas density there increases and a burst of star formation occurs,
as shown in Mayer et al. (2001b). This explains the bursty star
formation histories of Carina and Fornax.
Ram pressure 
and tidal stripping are reduced significantly at more recent epochs for
two reasons; the first is that in $\Lambda$CDM models orbits of satellites have on
average larger pericenters and longer orbital times at lower redshifts (see the 
cosmological simulations of Ghigna et al. (1998) for example), the second
is that both the local and cosmic UV background fade by orders of magnitude
at $z < 2$. 

In summary
we predict that dSphs with extended or truncated star formation histories
come from similar progenitors that fell into the primary halos at different
epochs. One variable whose evolution with time is not well known is the
density of the hot gaseous corona of the Milky Way which sets in part the
strength of ram pressure. However new simulations
of galaxy formation currently in progress suggest that the density of such
halo was higher at higher redshift, which would reinforce our scenario.


\begin{acknowledgments}
I thank Stelios Kazantzidis, Chiara Mastropietro, Ben Moore and Fabio
Governato for the many stimulating discussions on dwarf galaxies and 
Cristiano Porciani for his help on issues regarding the effects of the 
ionizing radiation.

\end{acknowledgments}

\end{document}